\documentclass[sigconf]{acmart}

\usepackage{subcaption}
\AtBeginDocument{%
  \providecommand\BibTeX{{%
    \normalfont B\kern-0.5em{\scshape i\kern-0.25em b}\kern-0.8em\TeX}}}

\setcopyright{acmcopyright}
\copyrightyear{2022}
\acmYear{2022}
\acmDOI{XXXXXXX.XXXXXXX}

\acmConference[Conference acronym 'XX]{Make sure to enter the correct
  conference title from your rights confirmation emai}{June 03--05,
  2018}{Woodstock, NY}
\acmPrice{15.00}
\acmISBN{978-1-4503-XXXX-X/18/06}

\begin{document}

\title{ECGT2T: Towards Synthesizing Twelve-Lead Electrocardiograms from Two Asynchronous Leads}

\author{Yong-Yeon Jo}
\email{yy.jo@medicalai.com}
\affiliation{\institution{Medical AI Inc.}
  \city{Seoul}
  \country{South Korea}
}
\author{Young Sang Choi}
\email{ychoi@ncc.re.kr}
\affiliation{
  \institution{National Cancer Center}
  \city{Goyang-si}
  \state{Gyeonggi-do}
  \country{South Korea}
}

\author{ Jong-Hwan Jang}
\email{jangood1122@medicalai.com}
\affiliation{\institution{Medical AI Inc.}
  \city{Seoul}
  \country{South Korea}
}

\author{Joon-myoung Kwon}
\email{cto@medicalai.com}
\affiliation{\institution{Medical AI Inc.}
  \city{Seoul}
  \country{South Korea}
}


\begin{abstract}
An electrocardiogram (ECG) is a non-invasive measurement used to observe the condition of the heart and usually consists of 12 synchronous leads. Wearable devices can record ECGs, but typically provide only a single lead or a couple of asynchronous leads. Signals generated from these devices may be insufficient for accurately diagnosing more complex cardiac conditions.
To bridge this gap, we propose \textsf{ECGT2T}, a deep generative model that synthesizes ten leads from two input leads to simulate a 12-lead ECG. 
Synthesized ECGs had timing and amplitude errors under 15 milliseconds and 10\%, respectively. 
Experiments on two widely used ECG datasets show classifiers trained on simulated 12-lead ECGs generated with ECGT2T outperformed models trained on one- or two-lead ECGs in detecting myocardial infarction and arrhythmia.
\end{abstract}





\maketitle

\section{Introduction}
An electrocardiogram (ECG) is a common and non-invasive way to help diagnose many cardiovascular conditions. There is increasing interest in using deep learning-based methods for ECG analysis, with recent studies proposing deep learning-based models for detecting atrial fibrillation~\cite{han19}, myocardial infarction~\cite{bal19}, and hypertrophic cardiomyopathy~\cite{ko20}. Parallel to these computational advancements, consumer wearable devices are becoming increasingly ubiquitous and can continuously record ECGs outside of the hospital setting. Due to physical and energy constraints, these devices usually have only one or two lead sensors in contrast to the 12 used in standard ECGs. A nonstandard number of leads can be insufficient for diagnosing more complex cardiovascular conditions and degrade the performance of deep learning-based diagnostic support systems~\cite{cho20}.


As a lead measures heart health along a particular axis, a standard 12-lead ECG can be generated by using projections from two asynchronous leads. To this end, we propose \textsf{ECGT2T}, a deep generative model for ECG synthesis that takes in two input leads and outputs ten leads to simulate a 12-lead electrocardiogram. We draw inspiration from generative adversarial network (GAN) based image-to-image translation models~\cite{hua17,kar2019}. Similar to how these networks generate images based off reference styles, ECGT2T learns lead styles during training, represents the cardiac condition from two input leads, and generates ten leads with a corresponding style. We quantitatively evaluate the generated signals by comparing their R-peaks, one of the important features used to diagnose heart disease, with the corresponding points in the original data. Additionally, we assess the quality of the ECGT2T outputs by visualizing the overlapping original and generated signals and confirm the applicability of the synthesized ECGs on downstream classification tasks. To demonstrate the importance of having a second asynchronous lead, we repeat the experiments with ECGS2E, a model with a similar architecture as ECGT2T with a single input lead and 11 generated lead outputs.

\section{Proposed Architecture}
\begin{figure*}[h]
\centering
\includegraphics[width=0.75\textwidth]{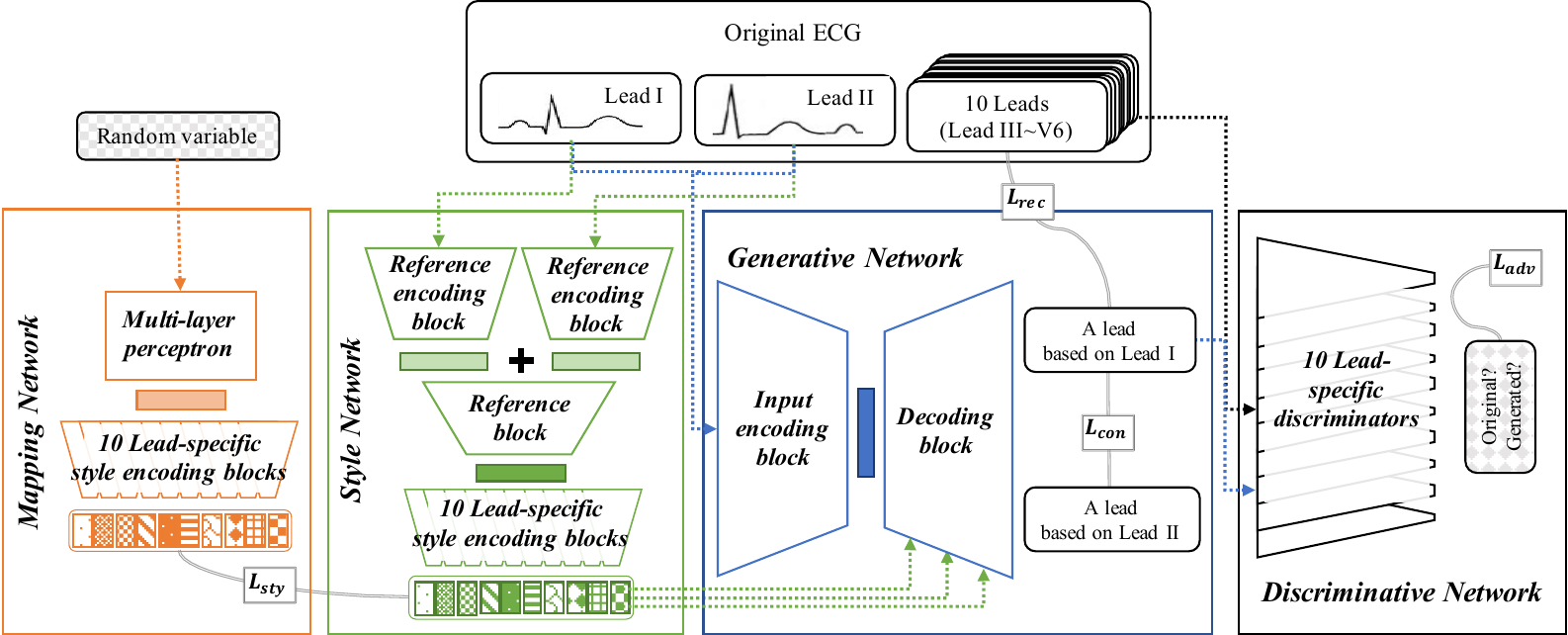}
\caption{ECGT2T model architecture. The model is comprised of style, mapping, generative, and discriminative networks. Each network is built with residual blocks. $L_{adv}$ is the adversarial objective, $L_{rec}$ is the reconstruction objective, $L_{con}$ is the lead consistency objective, and $L_{sty}$ is the style consistency objective.}
\label{fig:arch}
\end{figure*}

Figure~\ref{fig:arch} illustrates the ECGT2T architecture. The model consists of \textit{style} ($S(\cdot)$), \textit{mapping} ($M(\cdot)$), \textit{generative} ($G(\cdot)$), and \textit{discriminative} ($D(\cdot)$) networks. The style network represents the cardiac styles for ten output leads based on the styles of the two input leads. The mapping network, as seen in previous image style transfer models~\cite{kar2019}, generates latent codes for a random variable. The generative network takes any single lead, generates its respective latent code, and reconstructs the leads with a cardiac style. To improve the efficiency of this network, we use an adaptive instance normalization layer~\cite{hua17}. The discriminative network distinguishes whether its inputs are real or not. Each network is built by stacking multiple residual blocks~\cite{he16}.



%


\subsection{Objectives}

The full objective for ECGT2T is expressed as follows :
\begin{align*}
    \min_{E,H,G} \max_{D} \lambda_{adv} L_{adv} + \lambda_{rec} L_{rec}
    + \lambda_{con} L_{con} + \lambda_{sty} L_{sty} 
\end{align*}
where $\lambda$ and $L$ are the are the hyperparameters and losses for the respective objective.

The \textit{adversarial} objective is the mean of the log probability for original leads $D(x_{i})$ and the mean of the log of the inverted probabilities of the generated samples $D(G(x_{I},c_{i}))$:
\begin{align*}
    L_{adv} = \mathbb{E}_{x_{i}}[\log{D(x_{i})}] + \mathbb{E}_{x_{I},c_{i}}[\log{(1-D(G(x_{I},c_{i})))}]
\end{align*}
where $i$ is randomly selected lead that isn't Lead I or Lead II.

The \textit{reconstruction} objective compares an alternatively chosen Lead I or Lead II with a randomly selected lead $j$ via mean squared error:
\begin{align*}
    L_{rec} = \mathbb{E}_{x_{i},x_{j},c_{j}}[(x_{j} - G(x_{i},c_{j}))^2]
\end{align*}

The \textit{lead consistency} objective ensures that generated leads are consistent regardless of whether Lead I or Lead II was used as input.  For any randomly selected lead $i$ from Lead III to V6, the lead consistency objective is the mean squared error of the generated Lead I $G(x_{I},c_{i})$ and generated Lead II $G(x_{II},c_{i})$:
\begin{align*}
    L_{con} = \mathbb{E}_{x_{I},x_{II},c_{i}}[(G(x_{I},c_{i}) - G(x_{II},c_{i}))^2]
\end{align*}

To account for the diversity in  cardiac rhythm and beat shapes, we include the \textit{style consistency} objective as a form of regularization. This objective is the mean absolute error of the derived style $S(x_{I,II}, i)$ and mapping network output $M(z, i)$:
\begin{align*}
    L_{sty} = \mathbb{E}_{i,z,x_{I,II}}[||(M(z, i) - S(x_{I,II}, i))||_{1}]
\end{align*}
where $z$ is a random variable and $i$ is randomly selected from Lead III to V6.

\subsection{Training Procedure}
Adam optimization~\cite{kin14} with the learning rate of style, mapping, generative, and discriminative networks set to $3e^{-4}$, $1e^{-4}$, $3e^{-4}$, and $1e^{-4}$, respectively, and weight decay for all networks set to $1e^{-4}$. $\lambda_{rec}$ is set to 2, while the $\lambda$s for the other objectives are set to 1. All style latent vectors were set to size 512. The models used in this study are implemented with PyTorch and executed on a server equipped with Intel Xeon Silver 4210, 256 GB memory, and four NVIDIA RTX 3090 GPUs with 24 GB VRAM. We trained generative models (ECGT2T and ECGS2E) over seven days, and selected the model with the lowest loss.

\begin{figure*}[h]
\centering
\begin{subfigure}{.49\textwidth}
    \includegraphics[width=0.99\textwidth]{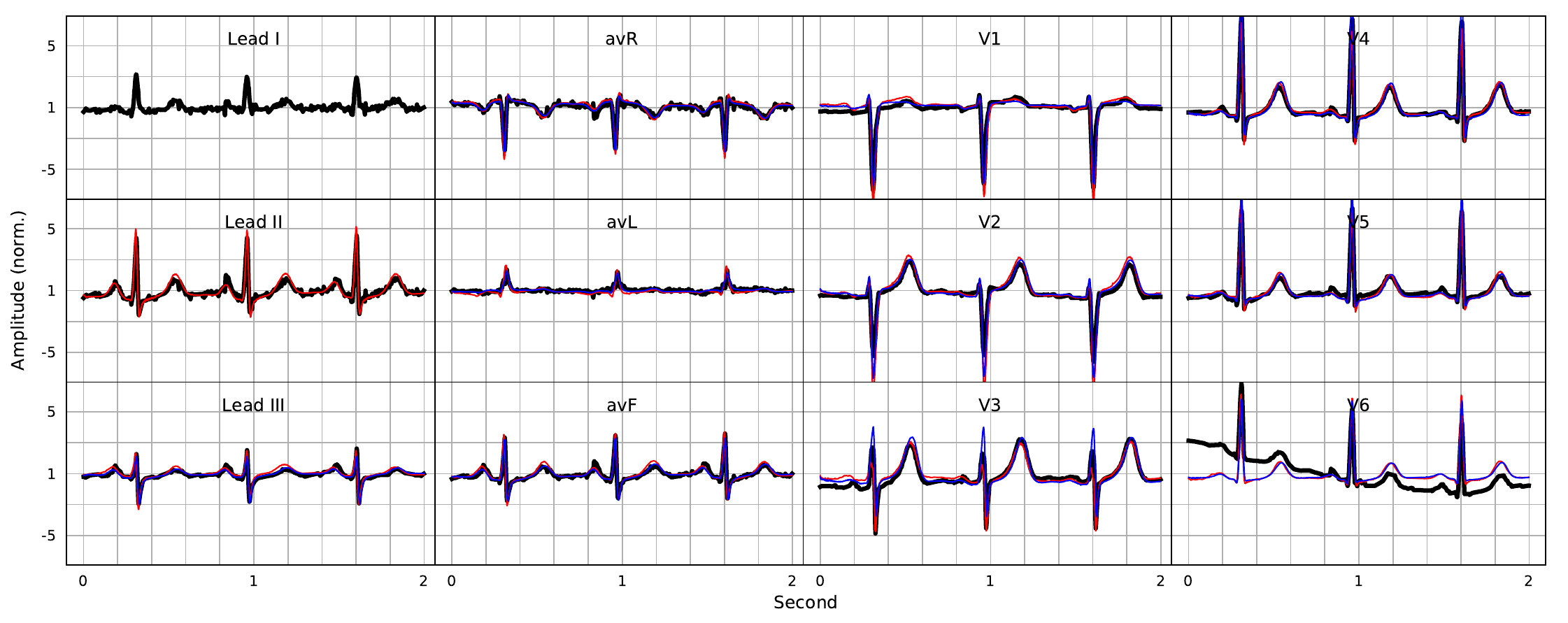}
    
    \caption{Normal sample from the PTB-XL dataset}\label{fig:ptb-norm}
\end{subfigure} 
\begin{subfigure}{.49\textwidth}
    \includegraphics[width=0.99\textwidth]{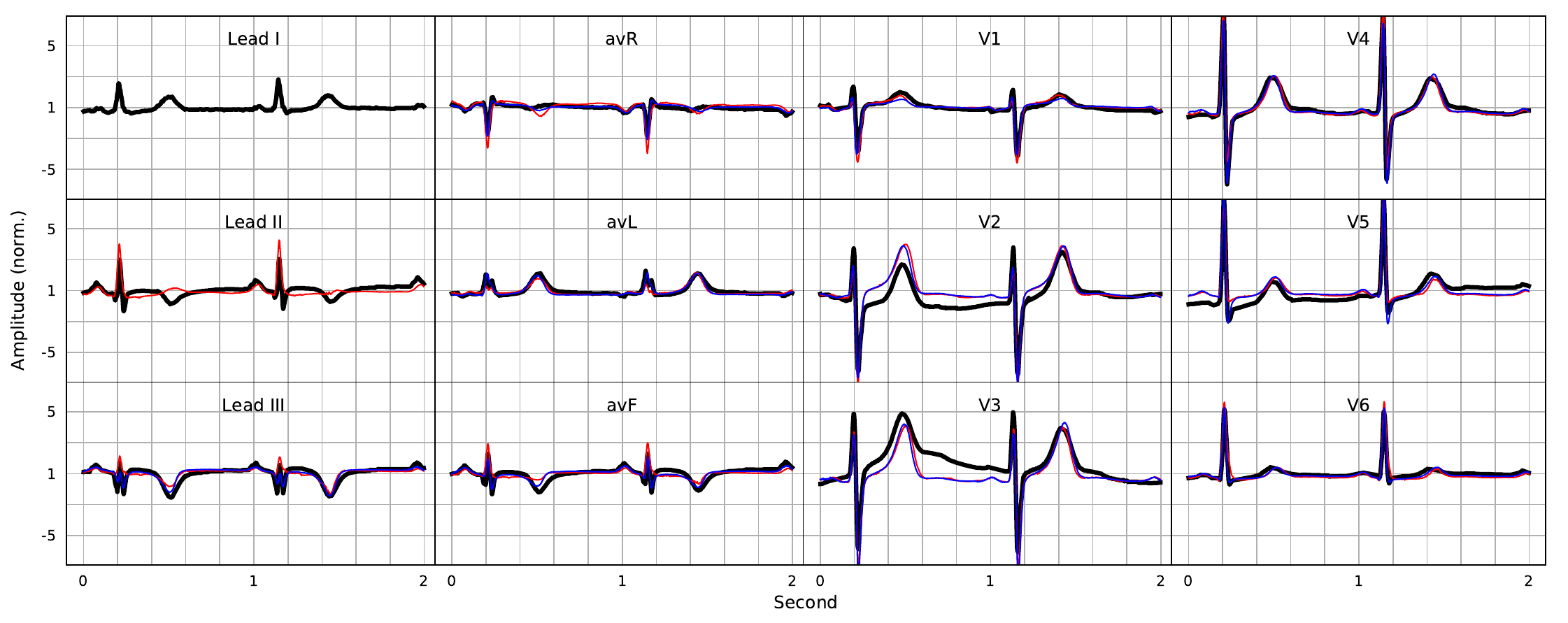}
    
    \caption{Myocardial infarction sample from the PTB-XL dataset}\label{fig:ptb-case}
\end{subfigure} \\
\begin{subfigure}{.49\textwidth}
    \includegraphics[width=0.99\textwidth]{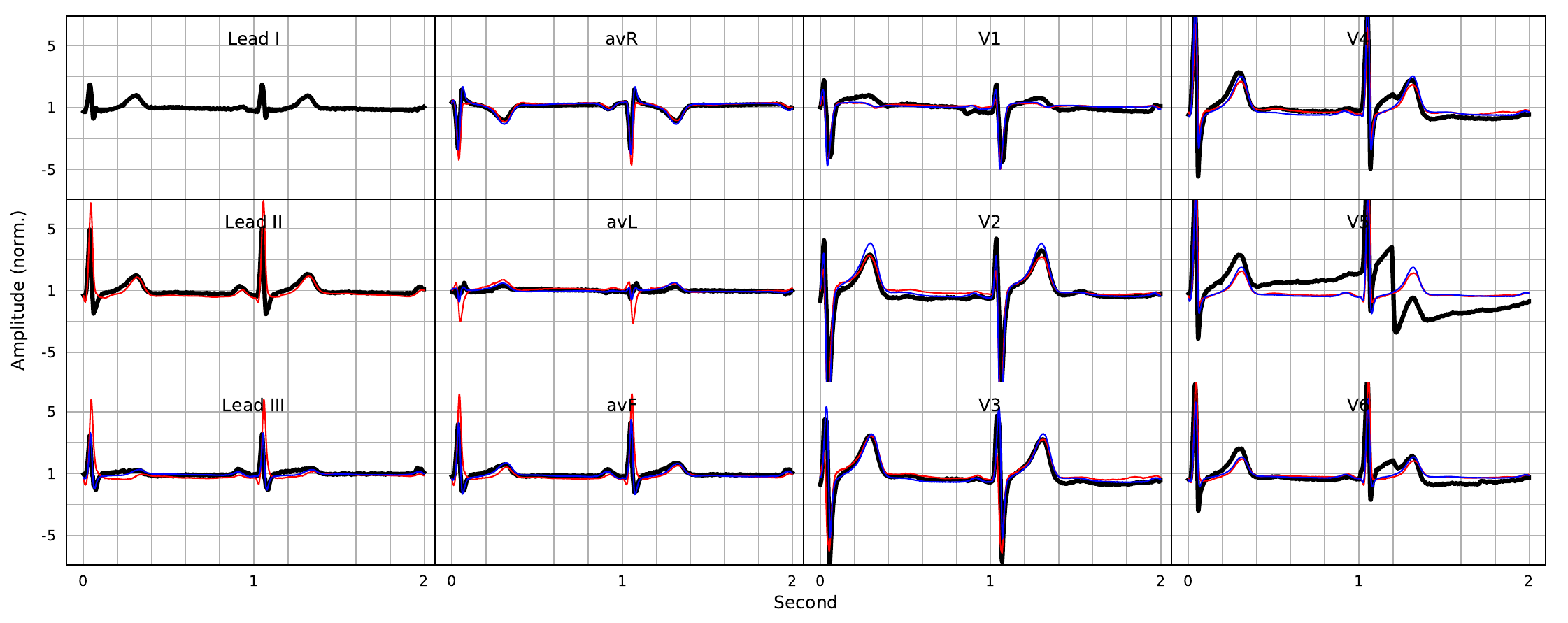}
    
    \caption{Normal sample from the CUSPH dataset}\label{fig:chap-norm}
\end{subfigure}
\begin{subfigure}{.49\textwidth}
    \includegraphics[width=0.99\textwidth]{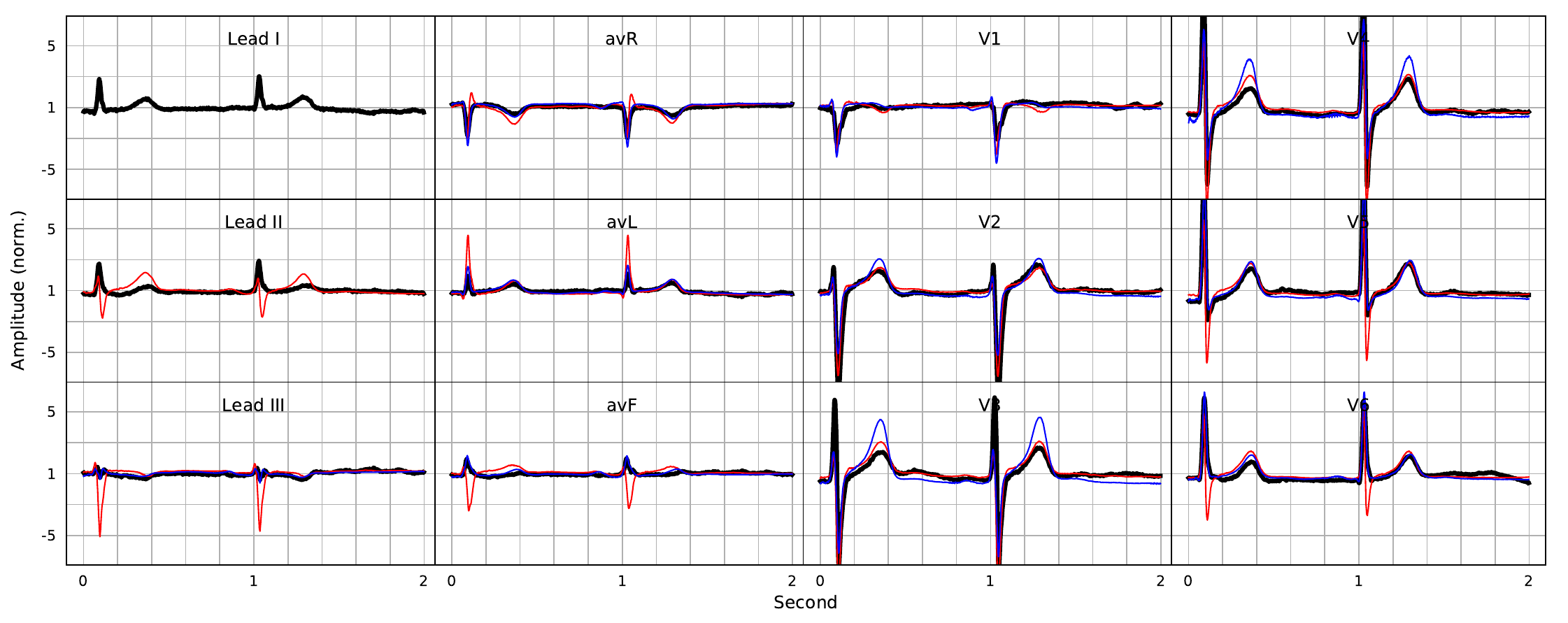}
    \caption{Atrial fibrillation sample from the CUSPH dataset}\label{fig:chap-case}
\end{subfigure}
\caption{12-lead electrocardiogram samples over a two second window. For each subplot, the black line denotes the original signal while the \textcolor{blue}{blue} and \textcolor{red}{red} lines represent the signals generated by \textcolor{blue}{ECGT2T}, and \textcolor{red}{ECGS2E} respectively. ECGS2E takes in only Lead I as input and outputs 11 leads, while ECGT2T takes asynchronous Lead I and Lead II inputs and generates signals for 10 leads. Although ECGT2T uses two asynchronous leads, all leads are visualized synchronously for convenience.
}
\label{fig:gen-ptb}
\end{figure*}

\section{Data and Experiments}
We use two widely used ECG datasets collected at the Physikalisch-Technische Bundesanstalt (PTB-XL)~\cite{wag20} and Chapman University and Shaoxing People's Hospital (CUSPH)~\cite{zhe20} for our experiments. We selected 15,012 ECGs (9,527 normal, 5,485 samples with any myocardial infarction) from the PTB-XL dataset and 9,094 (7,314 normal, 1,780 atrial fibrillation) samples from the CUSPH dataset. Both datasets were split into stratified training, validation, and test sets with a ratio of 7:1:2. ECGT2T was trained using only training samples from the PTB-XL dataset. 
The two asynchronous input leads were sampled by taking a signal in Lead I and then selecting the delayed signal 0.5 seconds later in Lead II. 

\subsection{Quality Assessment}

There are numerous ways to measure the discrepancy between two signals; however, common metrics such as root mean square error may not reflect the quality of generated electrocardiograms. If the original ECG waveform has artifacts such as power line interference, myokymia, and baseline wandering, a generated signal that accurately represents the cardiac condition will perform poorly under conventional metrics. Thus, we evaluated the quality of the ECGT2T outputs by comparing wave points between the original and generated signals.

An electrocardiogram has numerous points of clinical importance, and depending on a given patient's cardiac condition, some of these characteristics may not be visible in their electrocardiogram. However, the R-peak is both almost always present and is used for tasks such as  detecting heart rate and measuring stress level~\cite{sad12}. Therefore, we base our quality assessment with amplitude and positional missing errors for the R-peaks. We run our evaluation on V1 and V5 because of their electrode placement relative to Lead I: V1 is recorded orthogonally, while V5 is recorded in a similar position. We use Neurokit2~\cite{neu21} for R-peak detection on both the original and generated V1 and V5 leads.

Table~\ref{tab:gen_qual} shows the amplitude gaps and the position missing errors for the V1 and V5 leads generated by ECGT2T and ECGS2E for both datasets. For the PTB-XL dataset, the difference in amplitude is between 6.4$\sim$7.3\% and the positional missing error under 10 ms. For the CUSPH dataset the amplitude errors were higher than those in the PTB-XL dataset (7.6$\sim$11.4\% for amplitude and 3.8$\sim$14.4ms for position). We note that both models were only trained with samples from PTB-XL, which may explain this discrepancy.

Figure~\ref{fig:gen-ptb} shows 12-lead ECG samples over a two second window, where the black, red, and blue lines are the signals from the original electrocardiogram, ECGS2E, and ECGT2T respectively. Figure~\ref{fig:ptb-norm} and~\ref{fig:chap-norm} are normal samples from PTB-XL and CUSPH. The generated signals have less noise (seen in aVR and aVL in~\ref{fig:ptb-norm} and V4, V5, and V6 in~\ref{fig:chap-norm}) and baseline wandering (seen in V1, V3, and V6 in~\ref{fig:ptb-norm}). This artifact removal contributes to the amplitude gaps between the original and generated data seen in our quantitative evaluation.


Figure~\ref{fig:ptb-case} is a sample with myocardial infarction. ST-segment elevation, T-wave inversion, and abnormal Q-waves are characteristic of the condition and are captured by the generated outputs at Lead II, Lead III, V2, and V3 in this sample with the exception of the ST-segment elevation at the first beat in V3 for both models and the T-Wave inversion in Lead II for ECGS2E. The discrepancy between the ECGT2T and ECGS2E Lead II results is because Lead I and Lead II typically have peaks that point toward the same direction in normal ECGs. However, as ECGS2E takes in only Lead I as input, it initially misses the inversion. In contrast, ECGT2T has two inputs and is therefore unaffected by the discrepancy between Lead I and Lead II.

Figure~\ref{fig:chap-case} is a sample with atrial fibrillation. The condition is characterized by the absence of the P-wave, irregular ventricular rate, and slightly aberrant QRS complexes. In this case, all leads in the original ECG are missing the P-wave, and both ECGT2T and ECGS2E outputs capture this absence across all generated leads. However, while the ECGS2E limb leads have good placement, the amplitude of peaks  stray from the original signal. 

\begin{table}
  \caption{Synthetic quality assessment for the generated R-peaks. \textit{Amp} denotes the amplitude error while \textit{Pos} is the position error.}
  \label{tab:gen_qual}
  \begin{tabular}{cccccc}
    \toprule
    
        {Dataset} & Lead & \multicolumn{2}{c}{V1} & \multicolumn{2}{c}{V5} \\
                            \cmidrule{2-6}
                            & Model & ECGT2T & ECGS2E & ECGT2T & ECGS2E \\ 
        \midrule
        {PTB-XL} & Amp    & 6.4\%      & 6.4\%     &         7.3\%             &      6.7\%                  \\
                                & Pos     & 8.3ms      & 8.2ms     &        2.0ms   &  2.2ms                    \\ 
        \midrule
        {CUSPH}  & Amp    & 11.4\%      & 11.2\%     &     7.6\%&   7.7\%                   \\
                                & Pos     & 12.9ms      & 14.4ms     & 3.8ms   &  4.1ms \\

  \bottomrule
\end{tabular}
\end{table}

\subsection{Classification Performance}

To assess how signals generated by ECGT2T perform in downstream classification tasks, we configure six ECG datasets with different lead combinations: \textit{Original} is the baseline 12-lead ECG, \textit{T2T} is original asynchronous Lead I \& Lead II and ten generated leads, \textit{S2E} is the original Lead I and 11 generated leads, \textit{Two Leads} is composed of asynchronous Lead I \& Lead II only, and \textit{Single Lead} is the set with only Lead I. We train and test six classifiers for the corresponding ECG sets on myocardial infarction and arrhythmia detection tasks. 

We run our classification experiments on a customized ResNet18~\cite{he16} model. The classifier uses residual blocks with one-dimensional convolutions and a single fully connected layer for the output layer, and is trained with focal loss~\cite{lin17} with $\alpha$ and $\gamma$ set to 0.5 and 2, and Adam optimization~\cite{kin14} with learning rate and weight decay set to $1e^{-4}$ and $1e^{-5}$ respectively.

Table~\ref{tab:clf-perform} shows the classification performance by dataset. For myocardial infarction on the PTB-XL dataset, the model using \textit{T2T} outperforms the other models with the exception of the classifier using the original 12-lead electrocardiogram. Additionally, the performance of the model trained on the original Lead I and 11 generated leads (\textit{S2E}) is worse than those of the models using the original asynchronous Lead I and Lead II with no generated leads (\textit{Two Leads}). For atrial fibrillation detection on the CUSPH dataset, the classifiers trained with generated data outperform the models trained with a single lead or two leads. However, the S2E model outperforms T2T classifier, and the spread of performance metric results is much narrower than for the myocardial infarction task. This can be explained in part due to the differences in the two medical conditions: myocardial infarction is detected by the shape change of the heartbeat (which requires multiple leads), while arrhythmia can be diagnosed if there are irregular rhythmic patterns in the waveform, a symptom that can be detected with a single lead.

\begin{table}
\caption{Classification performance. First column indicates the training data source: \textit{Original} is the standard 12-lead ECG, \textit{T2T} is the original asynchronous Lead I \& Lead II with 10 generated leads, \textit{S2E} is the original Lead I with 11 generated leads, \textit{Two Leads} is the original asynchronous Lead I and Lead II with no additional generated leads, and \textit{Single Lead} the original Lead I only.  Parenthesis denote 95\% confidence intervals.}
  \label{tab:clf-perform}
  \begin{tabular}{ccccc}
    \toprule

\multicolumn{1}{c}{} & \multicolumn{2}{c}{Myocardial infarction} & \multicolumn{2}{c}{Arrhythmia}     \\
\midrule
          & AUROC & AUPRC & AUROC & AUPRC \\
         \midrule
Original &
\begin{tabular}[c]{@{}c@{}}
0.960\\(0.95-0.97)\end{tabular} &
\begin{tabular}[c]{@{}c@{}}
0.957\\(0.95-0.96)\end{tabular} &
\begin{tabular}[c]{@{}c@{}}
0.994\\(0.98-0.99)\end{tabular} &
\begin{tabular}[c]{@{}c@{}}
0.972\\(0.96-0.98)\end{tabular} \\
\midrule

T2T &
\begin{tabular}[c]{@{}c@{}}
0.948\\(0.94-0.96)\end{tabular} &
\begin{tabular}[c]{@{}c@{}}
0.945\\(0.94-0.95)\end{tabular} &
\begin{tabular}[c]{@{}c@{}}
0.990\\(0.98-0.99)\end{tabular} &
\begin{tabular}[c]{@{}c@{}}
0.960\\(0.94-0.97)\end{tabular} \\
\midrule

S2E &
\begin{tabular}[c]{@{}c@{}}
0.929\\(0.92-0.94)\end{tabular} &
\begin{tabular}[c]{@{}c@{}}
0.931\\(0.92-0.94)\end{tabular} &
\begin{tabular}[c]{@{}c@{}}
0.993\\(0.99-1.0)\end{tabular} &
\begin{tabular}[c]{@{}c@{}}
0.965\\(0.95-0.98)\end{tabular} \\
\midrule


Two Leads &
\begin{tabular}[c]{@{}c@{}}
0.937\\(0.93-0.95)\end{tabular} &
\begin{tabular}[c]{@{}c@{}}
0.938\\(0.93-0.95)\end{tabular} &
\begin{tabular}[c]{@{}c@{}}
0.989\\(0.97-0.99)\end{tabular} &
\begin{tabular}[c]{@{}c@{}}
0.954\\(0.93-0.96)\end{tabular} \\
\midrule

Single Lead &
\begin{tabular}[c]{@{}c@{}}
0.883\\(0.87-0.89)\end{tabular} &
\begin{tabular}[c]{@{}c@{}}
0.891\\(0.88-0.90)\end{tabular} &
\begin{tabular}[c]{@{}c@{}}
0.989\\(0.98-0.99)\end{tabular} &
\begin{tabular}[c]{@{}c@{}}
0.945\\(0.92-0.96)\end{tabular} \\

  \bottomrule
\end{tabular}
\end{table}


\section{Conclusion}
Modern wearable devices can record electrocardiograms with a couple of leads and collect increasingly large volumes of physiological signal data. However, the reduced number of leads used in these devices can be less accurate in measuring heart health or insufficient in detecting more complex cardiovascular conditions compared to standard 12-lead ECGs. To address these limitations, we propose ECGT2T, a deep generative model that simulates a standard 12-lead ECG from an input of two asynchronous leads by generating ten leads. We quantitatively evaluated the quality of the generated signals by comparing the R-peaks with those from the original ECG. Qualitatively, ECGT2T removed artifacts such as noise and baseline wandering that can impact clinical interpretation. For classifier performance, the models trained on generated signals outperformed models trained on one- and two- signal models in detecting myocardial infarction and arrhythmia.


\bibliographystyle{ACM-Reference-Format}
\bibliography{t2t}


\end{document}